\documentclass[preprint,12pt, 3p]{elsarticle}
\usepackage{graphicx}
\usepackage{latexsym}
\usepackage[utf8]{inputenc}
\usepackage{amsthm}
\usepackage[ruled,algonl,vlined, algo2e]{algorithm2e}

\usepackage{mathtools}  
\usepackage{amsmath}
\usepackage{algorithm}
\usepackage{algcompatible}
\usepackage{algpseudocode}

\usepackage{amssymb}
\usepackage{multirow}
\usepackage{url}
\usepackage{float}
\usepackage{xcolor}
\usepackage{booktabs}
\usepackage{fixltx2e}
\usepackage{adjustbox}
\usepackage{soul}
\usepackage{hyperref}

\usepackage{lineno}

\newcommand{\VK}[1]{{#1}}

\usepackage{caption,subcaption}
\usepackage{setspace}
\usepackage{fixmath}

\newtheorem{Example}{Example}
\newtheorem{Lemma}{Lemma}

\definecolor{newcolor}{rgb}{.8,.349,.1}

\journal{}

\begin{document}
\begin{frontmatter}
	\title{Conformal Group Recommender System}

	\author[nit]{Venkateswara Rao Kagita}
	\ead{venkat.kagita@nitw.ac.in}
 \author[nit]{Anshuman Singh}
\ead{as861850@student.nitw.ac.in}
	\author[DU]{Vikas Kumar~\corref{cor1}}
	\ead{vikas007bca@gmail.com}
	
\cortext[cor1]{Corresponding author}
\author[nit]{Pavan Kalyan Reddy Neerudu}
	\ead{neerud_951963@student.nitw.ac.in }
 \author[MU]{Arun K. Pujari}
	\ead{arun.pujari@mahindrauniversity.edu.in}
  \author[HCU]{Rohit Kumar Bondugula}
	\ead{rohitbond@uohyd.ac.in}
	%
	\address[nit]{National Institute of Technology Warangal, India.}
     \address[DU]{University of Delhi, Delhi, India.}
      \address[MU]{Mahindra University, Hyderabad, India.}
      \address[HCU]{University of Hyderabad, Hyderabad, India.}
	
\begin{abstract}
Group recommender systems (GRS) are critical in discovering relevant items from a near-infinite inventory based on group preferences rather than individual preferences, like recommending a movie, restaurant, or tourist destination to a group of individuals. The traditional models of group recommendation are designed to act like a black box with a strict focus on improving recommendation accuracy, and most often, they place the onus on the users to interpret recommendations. In recent years,  the focus of Recommender Systems (RS) research has shifted away from merely improving recommendation accuracy towards value additions such as confidence and explanation. In this work, we propose a conformal prediction framework that provides a measure of confidence with prediction in conjunction with a group recommender system \VK{to augment the system-generated plain recommendations}. In the context of group recommender systems, we propose various nonconformity measures that play a vital role in the efficiency of the conformal framework. We also show that defined nonconformity satisfies the exchangeability property. Experimental results demonstrate the effectiveness of the proposed approach over several benchmark datasets. Furthermore, our proposed approach also satisfies validity and efficiency properties.
\end{abstract}


\end{frontmatter}


\section{Introduction}
Recommender systems (RS) 
assist users in decision-making by helping them \VK{sift} through a huge variety of offered products, such as movies, web pages, articles, and books~\cite{ricci11}. These systems exploit past interactions between users and items to make personalized recommendations. RS algorithms are broadly classified into content-based, collaborative, and hybrid filtering, \VK{depending on the input used for profiling users and items.} Content-based filtering approaches recommend items to a user by considering the similarity between the content or features of items and the user's profile~\cite{pazzani2007content, lops2011content}. On the other hand, the collaborative filtering-based approach recommends items based on the preferences of other users who share similar preferences to the target user~\cite{koren2021advances, kumar2017collaborative, salman2016combining}. The hybrid approach combines various mechanisms and compositions of different data sources~\cite{burke2002hybrid, bostandjiev2012tasteweights}. Group recommender systems (GRS)~\cite{kagita2013group,kagita2013precedence} extend this concept to the user group, wherein it analyses a group of users’ profiles and creates a communal recommendation list. These groups can consist of family members, friends, colleagues, or any collection of individuals who wish to engage with a specific item or application collectively. We observe numerous applications of group recommender systems in our daily lives. For example, a group of friends may plan to dine at a restaurant or organize a tour, while a family may wish to watch a movie together. A Group Recommender System (GRS) can be formally defined as follows. Let $O = \{o_1, o_2, \ldots, o_m\}$ represent the set of items, $U = \{u_1, u_2, \ldots, u_n\}$ denote the set of users, $O_j$ represent the set of items present in the profile of a user $u_j$, and $U_G\subseteq U$ indicate a group of users. Given the sets, $O$, $U$, $O_j$ ($\forall j$), and $U_G$, the goal of a GRS is to recommend the top-k most relevant items from $O$ to the user group $U_G$ by considering the individual preferences within the group.

Several approaches in recent years have been proposed to extend personalized recommender systems to group recommender systems, wherein the major focus is to improve the accuracy of recommendations.  With the recent algorithmic advancement, the focus of the research has been recently shifted towards 
creating transparent group recommendation models that prioritize accountability and explainability. These models aim to achieve accuracy while providing additional value through confidence measures, explanations, or sensitivity. Among these enhancements, associating a confidence measure with the recommendation set is a particularly appealing facet. The confidence measures indicate the system's confidence that the desirable items are present in the recommendation set, which in turn enhance the system's reliability and help users quickly and correctly identify the products of their choice. Although researchers have thoroughly investigated confidence-based personalized recommender systems~\cite{kagita2017conformal, kagita2022inductive}, only a few methods have been proposed for improving the confidence measure in group recommender systems. Further, considering the individual preferences of the group and providing a delightful recommendation with confidence is not trivial. This paper proposes a confidence-based group recommender system using a conformal prediction framework. The proposed approach represents confidence in connection with the error bound, i.e., if the system exhibits $80\%$ confidence in the recommendation, it implies that the probability of making an error is at most $20\%$. The concept of conformal prediction forms the basis for the proposed Conformal Group Recommender System (CGRS). 
  
Conformal prediction is a framework for reliable machine learning that measures confidence in the predicted labels on a per-instance basis. A conformal system framework can be implemented alongside any traditional machine learning algorithm, but the implementation largely depends on the underlying algorithm. This paper extends the concept to GRS and defines a nonconformity measure, an essential part of the conformal prediction framework, suitable for group recommendation settings. Given a set of users $U$, items $O$, and a target group of user profiles, $O_j (\forall j),~ U_G$ and the significance level $\varepsilon$, the proposed conformal group recommender system returns recommendation set to a group $U_G$ with the confidence of $(1-\varepsilon)$. We demonstrate that the exchangeability property, an essential property for any conformal framework, is satisfied by the proposed nonconformity measure. Furthermore, the proposed approach also satisfies the validity and efficiency properties. Experiment results over several real-world datasets support the efficacy of the proposed conformal group recommender system. 


The paper is organized as follows. Section~\ref{sec:RW} briefly reviews conformity measures in recommender systems and establishes the context for this paper. We cover the foundational concepts of conformal prediction and the algorithm used to build the proposed conformal group recommender systems in Section~\ref{sec:FC}. The proposed algorithm and proof of validity is presented in Section~\ref{sec:CGRS}. Section~\ref{sec:Exp} presents extensive experimental results on eight benchmark datasets. Finally, Section~\ref{sec:ConFw} concludes the paper and offers recommendations for future endeavors.


\section{Related Work}
\label{sec:RW}

Group recommender systems assist a group of users who share a common interest. These systems exploit the collective consumption experience of the group to offer them personalized recommendations. However, most of these systems are less transparent than those designed to assist a single user, making it difficult for users to understand why the system recommends a particular item. Despite the enormous application of group recommender systems, there has yet to be a comprehensive analysis of associating confidence with the recommendation in a group context. While confidence measures are well-established in personalized recommender systems~\cite{kagita17,kagita2022inductive}, their application in group recommender systems is more complex. Group recommender systems have additional complexities in modeling and evaluating preferences due to diverse user preferences. 
This section reviews related work that focuses on group recommender systems and associating confidence measures with the recommendation set.

The majority of conventional group recommendation models aim to extend personal recommender systems (PRS) to group recommender systems (GRS) by blending individual preferences to form group profiles. MusicFX ~\cite{mccarthy1998musicfx}  is among the early proposals in group recommendation, specifically targeting the recommendation of music channels for playing in fitness centers.  This system exploits previously specified music preferences of individuals over a wide range of musical genres to recommend music at a particular time. Let's browse~\cite{lieberman1998let}  suggests web pages to a group consisting of two or more individuals who are likely to have shared interests.PolyLens~\cite{o2001polylens} employs collaborative filtering and recommends movies to groups of users. Jameson et al.~\cite{jameson2004more}  proposed the Travel Decision Forum system, which helps groups plan vacation. The system provides an interface where the group members can view and copy other users' preferences and combine them using the median strategy once all agree. Garcia et al.~\cite{garcia2009group} proposed intersection and aggregation mechanisms to build a group profile from the individual profiles. Kagita et al.\cite{kagita15} use a virtual user approach with precedence mining to extend PRS to GRS. All the GRS approaches described so far utilize aggregation techniques to determine group profiles or recommendations, and they are inept at associating confidence with the recommendation. In this proposal, we adopt our previous work~\cite{kagita15} and propose a conformal group recommender system that furnishes confidence in the recommendations to improve the system's reliability and user satisfaction. Another difference between the previous work~\cite{kagita15} and our current proposal is that we have used coexistence in place of precedence mining, as in many real-world applications, the associations of item consumption are more important than strict precedence relationships.

\section{Preliminary Concepts}
\label{sec:FC}
This section reviews the paper's foundational concepts, which we use subsequently in our algorithm. We introduce the precursory concepts of conformal prediction in Section~\ref{subsec:CP}, which serves as the framework for constructing our suggested conformal group recommender system.
We then briefly describe the precedence mining and association mining-based personalized recommender systems (PRS) in Sections~\ref{subsec:PM} and ~\ref{subsec:AM}, respectively, that we extended to the GRS. Section 3.4 explores the virtual user approach as an efficient method for developing a group recommender system (GRS) from a personalized recommender system (PRS).

\subsection{Conformal Prediction}
\label{subsec:CP}
In this section, we offer a concise overview of the conformal prediction principle, presenting relevant information for a comprehensive understanding.
Let $z_{i}= (x_{i}, y_{i})$ be a training instance (or an example), where  $x_{i} \in \mathbb{R}^{d}$ is a $d$-dimensional attribute vector and $y_{i} \in \{Y_1, Y_2, \dots, Y_c\}$ is the associated class label. Given a dataset, $S = \{z_{i}\}_{i=1}^n$, which contains $n$ training instances, the aim of the classification task is to produce a classifier $h: x \rightarrow y$, anticipating that $h$ predicts $y_{n+k}$ well on any future example $x_{n+k}$, $k\ge 1$, by optimizing some specific evaluation function. A conformal predictor, on the other hand, associates precise levels of confidence in predictions. Given a method for making a prediction, the conformal predictor offers a set of predicted class labels for an unseen object $x_{n+k}$  such that the probability of the correct label not present within this set is no greater than $\varepsilon$, called $(1-\varepsilon)-$prediction region.  We assume that the set of all possible labels  $\{Y_1, Y_2, \dots, Y_c\}$ is known \textit{a prior}, and all $z_{i}'s$, are in i.i.d. fashion (independently and identically distributed).  

The conformal predictor uses a (non)conformity measure designed based on the underlying algorithm to evaluate the difference between a new (test) instance and a training data set. Let $z_{n+1} = (x_{n+1}, Y_j)$, where  $Y_j$ is tentatively assigned to $x_{n+1}$. The nonconformity measure for an example $z_i \in \{S\cup z_{n+1}\}$ is a measure of how well  $z_i$ conforms to $\{S\cup z_{n+1}\} \setminus z_i, \forall i\in [1,n+1]$. 
In other words, the nonconformity measure assesses the likelihood that the label $Y_j$ is the correct label for our new example, $x_{n+1}$. The difference in $S$'s predicted behaviour when $z_i$ is replaced with $z_{n+1}$ is then calculated. Let $p$-value be the proportion of $z_{i}\in S$ with a nonconformity score worse than $z_{n+1}$ for all possible values of $y_{n+1}$ (all class labels). The $(1-\varepsilon)$-prediction region is formed by labels with $p$-values greater than $\varepsilon$. Intuitively, the prediction behavior is observed by employing any of the conventional predictors that utilize the training set $S$.

\subsection{Precedence Mining based Recommender Systems}
\label{subsec:PM}

The \emph{precedence mining (PM)} based recommendation model is a type of collaborative filtering (CF) technique that captures the temporal patterns in the user profiles and provides recommendations to the target user based on the sequence of items they have consumed~\cite{param}. For example, if a user has seen \textit{Toy Story 1}, the PM-based CF can recommend \textit{Toy Story 2} if it has been seen previously by users who have seen the \textit{Toy Story 1}. The PM model uses precedence statistics, i.e., the precedence count of item pairs, to estimate the probability of future consumption based on past behavior. It estimates the relevance score of each candidate item for recommendation using precedence statistics. 

Let $O= \{o_{1}, o_{2}, \ldots , o_{nitem}\}$ be the set of items, $U = \{u_{1}, u_{2}, \ldots,u_{nuser}\}$ be the set of users and $O_{j}$ be the set of items consumed by $u_j$.
The items for recommendation are chosen based on the criteria that they are not present in the target user's profile and are likely to be preferred by the target user compared to other items.
The PM-based approach keeps $Support(\cdot)$ and $PrecedenceCount(\cdot)$ statistics for each item $o_i$ to calculate the recommendation score of an item.  $Support(o_i)$ is the number of users who has consumed the $i$th item, and the $PrecedenceCount(o_i, o_h)$ is the number of users who consumed item $o_i$ before item $o_h$. We use a notation $PP(o_{i}|o_h)$ to denote the precedence probability of consuming an item $o_{i}$ preceding $o_{h}$. We define  $PP(o_{i} | o_{h})$, and $Score(o_{i},u_{j})$ as follows.

\begin{equation}
 PP(o_i|o_h) = \frac{PrecedenceCount(o_i, o_h)}{Support(o_h)},
\end{equation}

 \begin{equation}
  Score(o_i, u_j) =  \frac{Support(o_i)}{nuser} \times \underset{o_l \in
O_j}{\operatorname{\prod}}PP(o_l|o_i).
 \label{eq:score1}
 \end{equation}
 
\noindent The objects with high scores  are recommended.

\begin{Example}
\label{ex1}
    
    Let us consider the movie consumption data of nine users watching six different movies (i.e., Toy Story, Minions, Godfather I, Godfather II, Thor, and Avengers) in the table given below. Let \emph{Rob} be the target user who has not yet watched \emph{Minions} and \emph{Godfather II}. We compute the relevance score for these two candidate items and rank them in decreasing order of the relevance score. 
    
    \begin{table}[H]
    \centering
    \begin{tabular}{|l|l|} 
    \hline
    \textbf{Users} & \textbf{Movies watched}                        \\ 
    \hline
    Mark           & Minions, Thor, Toy Story, Avengers             \\ 
    \hline
    Christopher    & Godfather I, Godfather II, Toy Story, Minions  \\ 
    \hline
    Rob            & Thor, Avengers, Godfather I, Toy Story         \\ 
    \hline
    Jacob          & Avengers, Godfather I, Godfather II, Thor      \\ 
    \hline
    Rachel         & Godfather I, Minions, Godfather II, Toy Story  \\ 
    \hline
    Thomas         & Godfather I, Godfather II, Minions                \\ 
    \hline
    Grant          & Thor, Avengers, Minions, Toy Story             \\ 
    \hline
    Pamela         & Godfather I, Avengers, Thor, Godfather II       \\ 
    \hline
    Holly          & Minions, Toy Story, Godfather I, Godfather II  \\
    \hline
    \end{tabular}
    \end{table}
    
    \begin{align*}
    Score(Godfather II, Rob) &= \frac{Support(Godfather~II)}{nuser}\times PP(Thor\mid Godfather II)\times\\
    &~ ~  PP(Avengers\mid Godfather~II)\times PP(Godfather~I\mid Godfather~II)\\
    &~ ~ \times P(Toy~story\mid Godfather~II) \\
                  &= \frac{6}{9} \times \frac{1}{6}\times\frac{2}{6}\times \frac{6}{6}\times \frac{1}{6} = 0.0062. 
\end{align*}
\end{Example}

\subsection{Association Mining based Recommender Systems}
\label{subsec:AM}
Although the temporal pattern is important in some real-world applications, the associations of item consumption are more important than strict precedence relationships. For instance, in household product recommendations, device recommendations,  or Youtube video recommender systems, more than the sequence, it is crucial to analyze the set of items consumed together. Consider the table given in Example \ref{ex1} where some movies, such as \emph{Minions} and \emph{Toy Story}, are watched together but not necessarily in a particular order. Out of the six users who have watched the movie \textit{Minions}, five of them have also watched \textit{Toy story} either before or after it. The opposite is also true.  This could be because, while \textit{Minions} and \textit{Toy Story} are not as closely related as \textit{Godfather I} and \textit{II}, they are both animated children's films, and it is likely that if a person enjoys \textit{Minions}, he or she will enjoy \textit{Toy Story} as well. We consider the number of profiles that consumed two items (say, $o_i$ and $o_j$) together in this work rather than the number of profiles that consumed $o_2$ after $o_1$. In other words, instead of taking the precedence probability of $o_j$ given $o_i$ ($P(o_j\mid o_i)$), we take the probability of $o_i$ given $o_j$ ($P(o_i \mid o_j)$) while estimating the relevance score of an item to the target user. Thus, Equation~\ref{eq:score1} is rewritten as 

 \begin{equation}
  Score(o_i, u_j) =  \frac{Support(o_i)}{nuser} \times \underset{o_l \in
O_j}{\operatorname{\prod}}P(o_i|o_l),
\label{eq:score2}
 \end{equation}
where $P(o_i|o_h) = \frac{Support(o_i, o_h)}{Support(o_h)}$ and $Support(o_i, o_h)$ is the number of users having consumed item $o_{i}$ and $o_{h}$ together. $Support(o_i)$ represents the number of users who have consumed item $o_i$, which is also equal to $Support(o_i, o_i)$. $P(o_{i}|o_h)$ denotes the probability of consuming an item $o_{i}$ given that an item $o_h$ is already consumed. 
Items with high scores are subsequently recommended to a user. Conversely, if the score is low, it is implausible that the item would interest the user.

We call this formulation as \emph{Association Mining based Recommender System}.  We now consider an example which illustrates the working of the Association Mining based recommender system. 

\begin{Example}
     Take into account the Support statistics below, computed based on thirty users' profiles. $U =
\{u_1, u_2, \ldots, u_{30}\}$ over ten items $O = \{o_1, o_2, \ldots, o_{10}\}$.

\[
Support = \left[
\begin{array}{cccccccccc}
 20 & 9 & 8& 11&  7 & 8 & 6 & 7 & 7 & 3 \\
 9 & 25& 10 &11 & 9 & 7 & 7 & 6 & 8 & 4 \\
 8 & 10 & 21 & 5&  7 & 6 & 5 & 6 & 4 & 3 \\
 11 & 11 & 5 & 25 & 6 & 8 & 6 & 6 & 3 & 2 \\
 7 & 9 & 7 & 6 & 22 & 8 & 7 & 6 & 9 & 4 \\
 8 & 7 &6 & 8 & 8 & 18 & 5 & 6  & 4 & 1 \\
 6 & 7 & 5 & 6 & 7 & 5 & 15 & 4 & 4 & 1 \\
 7 & 6 & 6 & 6 & 6 & 6 & 4 & 18 & 7 & 2 \\
 7 & 8 & 4 & 3 & 9 & 4 & 4 & 7 & 20 & 3 \\
 3 & 4 & 3 & 2 & 4 & 1 & 1 & 2 & 3 & 6 \end{array} \right] \]

\noindent Let $u_1$ be the target user and $O_1 = \{o_1,~ o_3,~ o_5,~ o_7,~ o_9\}$ be set of items consumed by $u_1$. The items which are not consumed by the the user $u_1$ forms the candidate item set for recommendation i.e., $O\setminus O_1 = \{o_2, o_4, o_6, o_8, o_{10}\}$.
The score of an item $o_2$ which not consumed by user $u_1$ is then calculated as
\begin{align*}
    Score(o_2, u_1) &= \frac{Support(o_2)}{30}\times P(o_1\mid o_2)\times P(o_3\mid o_2)\times P(o_5\mid o_2)\times P(o_7\mid o_2) \times P(o_9\mid o_2)\\
                  &= \frac{25}{30} \times \frac{9}{25}\times\frac{10}{25}\times \frac{9}{25}\times \frac{7}{25}\times  \frac{8}{25} = 0.0039.
\end{align*}

\noindent Similarly,  $Score(o_4, u_1) = 0.0005$,  $Score(o_6, u_1) = 0.0024$, $Score(o_8, u_1) = 0.0022$, and
 $Score(o_{10}, u_1) = 0.0028$. Hence, it ranks the items in the order of $o_2,~ o_{10},~ o_6,~ o_{8}$, and $o_{4}$. 
 \end{Example}
 
 
 Recommender systems that employ Association Mining focus on discovering associations among items that users have consumed. These systems then utilize support statistics to compute a high \emph{relevance score} for recommending new items. The major advantage of the current model over a traditional CF approach is its ability to account for the  frequent pairwise relations among all users.  
 After identifying a subset of similar users, the traditional CF approach narrows down its search to items consumed by the users within the neighborhood. Consequently, certain consumption patterns of items by the entire user group are excluded due to this restriction. 
 The nicety and novelty of this approach lie in its utilization of pairwise relations between items. The current approach is also different from the association rule mining~\cite{kim2011recommender} that derives if-then rules using user transactions. In contrast, the present method estimates the relevance score as the likelihood of consuming a candidate item given a set of commodities consumed by a user.

\subsection{Virtual User Approach for GRS}
\label{subsec:VM}
The virtual user approach for GRS creates a virtual profile representing the group preferences and utilizes this profile to build the recommendation set~\cite{kagita15}. A virtual user profile is created by computing the weight that indicates the group preferences for each item consumed by at least one group member. We compute the weight of an item $o_i$ for a group $G$ as follows: 
\begin{equation}
\label{eq:weight}
    weight(o_i, G) = \sum_{u_j\in G} \frac{weight(o_i, u_j)}{\lvert G \rvert}~ ~ where, \\
 \end{equation}
\[  
    weight(o_i, u_j) = \begin{cases}
    1,& \text{if } o_i \in O_j  \\
    score(o_i, u_j),   & \text{otherwise}
\end{cases}.
\]

In our previous work, after determining the group weight for all the group items, we proposed two different ways of creating a virtual user profile: 1)\emph{Threshold-based virtual user $-$}  A virtual user profile is made up of items whose weight exceeds a certain threshold. 2) \emph{Weighted virtual user $-$} The virtual user profile comprises all the items each group member consumes and their weights. This paper adapts both of these techniques and proposes an \emph{hybrid approach} that relieves the items with significantly less weight and forms a weighted virtual user profile with the remaining items.

\section{Conformal Group Recommender System}
\label{sec:CGRS}
The proposed approach combines conformal prediction and group recommender systems to enrich the system-generated plain group recommendations with the associated confidence value. As discussed previously, the nonconformity measures, a pivotal component of the conformal prediction framework, play a critical role in the efficiency of any conformal model and must be devised carefully after inspecting the underlying model. The nonconformity measure, which essentially measures how strange the new object is in comparison to the training set, is then used to derive a confidence value for a new item. We use the pairwise relations between the items to derive the nonconformity measure in the proposed work. 


 \subsection{Nonconformity Measure for CGRS}
 \label{sec:nc_CGRS}
 Let $V(G)$ be the virtual user representing the group $G$ profile and $O_{V(G)}$
 be the set of weighted items in the virtual user profile.   We randomly divide the virtual user profile $O_{V(G)}$ into $O_{V(G)}^1 = \{o_1:w_1, o_2:w_2, \ldots, o_n:w_n\}$ and $O_{V(G)}^2  = \{o'_1, o'_2, \ldots, o'_k\}$, where $O_{V(G)}^1$ is the training set that holds items known to be consumed and $O_{V(G)}^2$ contains the candidate set of items that are consumed by the user but are not part of the training dataset. Given a new candidate item $o_{n+1} \in O_{V(G)}^2$,  CGRS measures its strangeness with respect to the items in the training set. We define the nonconformity of an item $o_i \in O_{V(G)}^1$
as the relevance score of a candidate item $o'_j \in O_{V(G)}^2$ with a profile excluding  $o_i$. We compute the nonconformity value for each item in the extended training set  $O_{V(G)}^{3}= O_{V(G)}^1 \cup \{o_{n+1}:1\}$. To compute the nonconformity score of an item $o_i$, we use the profile $O_{V(G)}^{4} = O_{V(G)}^{3}\setminus \{o_i:w_i\}$. The nonconformity value of $o_i$, indicated as $\alpha^i$, concerning the recommendability of an item $o'_j$ is defined as follows. 
\begin{equation}
\label{eq:ncm}
    \alpha^i(o'_j) =  \frac{Support(o'_j)}{nuser} \times \underset{o_l \in O_{V(G)}^{4}}{\operatorname{\prod}} w_l \times P(o_l|o'_j).
\end{equation}

\begin{Lemma}
\label{lem:invariant}
 Nonconformity $\mathcal{A} ()$ of items remains consistent regardless of the permutation $\pi$ of the items, i.e.,~$\mathcal{A}(o_{1}, o_{2}, \ldots, o_{n+1}) = (\alpha^{1}, \alpha^{2}, \ldots,  \alpha^{n+1})\Rightarrow  \mathcal{A}(o_{\pi(1)}, o_{\pi(2)},\\ \ldots,
		o_{\pi(n+1)}) = (\alpha^{\pi(1)}, \alpha^{\pi(2)}, \ldots, \alpha^{\pi(n+1)}).$
	\end{Lemma}
	\begin{proof}
  The nonconformity computation given in Equation~\ref{eq:ncm} considers the pairwise probability of a recommendable item $o'_j$ concerning all the items of $O^4_{V(G)}$ in the product term. One can easily see that the result is invariant of the order of the items in $O^4_{V(G)}$. Hence, the nonconformity scores remain unaffected even when the training set is permuted.
  Hence, it is proved that the proposed nonconformity measure yields consistent results regardless of the permutation of the items. 
	\end{proof}


\subsection{Recommendation using $p$- values}
\label{sec:pvalue}
The  $p$-value of a new item $o_{n+1}$ is defined as the proportion of $o_i$'s in $O_{V(G)}^{4}$ such that the score  $\alpha^i(o'_j)$ is greater than or equal to $\alpha^{n+1}(o'_j)$.
\begin{equation}
    p(o_{n+1}, o'_j) = \frac{\Big\lvert\{o_i\mid 1\le i\le n+1, \alpha^i(o'_j)\ge \alpha^{n+1}(o'_j)\}\Big\rvert}{n+1}
    \label{eq:p-value1}
\end{equation}

\noindent It can also be seen that the disagreement between $\alpha^h
(o'_j )$ and $\alpha^t (o'_j ))$ for $o_h \neq o_t$ is proportional to the difference between probabilities ${P}(o_t|o'_j)$ and $P(o_h|o'_j)$.  Formally, $\alpha^h (o'_j ) \ge \alpha^t(o'_j )$ if and only if $(w_h\times P(o_h|o'_j)) \le (w_t \times P(o_t|o'_j))$. 
\begin{Lemma}
        $\alpha^h (o'_j ) \ge \alpha^t(o'_j )$ if and only if $(w_h\times P(o_h|o'_j)) \le (w_t \times P(o_t|o'_j))$
\end{Lemma}
\begin{proof}
\begin{align*}
    \alpha^h (o'_j ) & =  \frac{Support(o'_j)}{nuser} \times \underset{o_l \in O_{V(G)}^{4}}{\operatorname{\prod}} w_l \times P(o_l|o'_j).\\
    & =  \frac{Support(o'_j)}{nuser} \times (w_1 \times P(o_1|o'_j)) \times\ldots \times  (w_{h-1} \times P(o_{h-1}|o'_j))\times (w_{h+1} \times P(o_{h+1}|o'_j))\times \\ 
    &~ ~ \ldots \times (w_{n+1} \times P(o_{n+1}|o'_j)) .
\end{align*}

\noindent Similarly, 
\begin{align*}
    \alpha^t (o'_j ) & =  \frac{Support(o'_j)}{nuser} \times (w_1 \times P(o_1|o'_j)) \times\ldots \times (w_{t-1} \times P(o_{t-1}|o'_j))\times (w_{t+1} \times P(o_{t+1}|o'_j))\times \\
    &~ ~ \ldots \times (w_{n+1} \times P(o_{n+1}|o'_j)) .
\end{align*}

\noindent The only term missing in $\alpha^h$ computation is $(w_{h} \times P(o_{h}|o'_j))$ and similarly, for $\alpha^t$ computation is $(w_{t} \times P(o_{t}|o'_j))$. The rest of the terms are the same in both $\alpha^h$ and $\alpha^t$ computation. Hence,  $\alpha^h (o'_j ) \ge \alpha^t(o'_j )$ if and only if $(w_h\times P(o_h|o'_j)) \le (w_t \times P(o_t|o'_j))$. 
\end{proof}

Hence, comparing the weighted probabilities is better choice than comparing the scores directly to streamline the computation. Thus we redefine  $p(o_{n+1},o'_j)$ given in Equation~\ref{eq:p-value1} as follows.


 \begin{equation}
 \label{eq:pval2}
     p(o_{n+1},o'_j)=  \frac{\Big\lvert\{ o_h\mid 1\le h \le n+1, w_h\times P(o_h|o'_j) \le w_{n+1}\times P(o_{n+1}|o'_j)\}\Big\rvert}{n+1}
 \end{equation}

\noindent 
The final $p$-value is then computed by taking an average of $p$-values concerning  each candidate item in the set $O^2_{V(G)}$.  A higher average $p$-value of an item implies a greater chance of being recommended.

\begin{equation}
\label{eq:pval3}
p(o_{n+1}) = \frac{\sum_{o'j\in O^2_{V(G)}} p(o_{n+1}, o'_j)}{k}
\end{equation}

\noindent \textit{Recommendation set $\Gamma^\varepsilon$ with $(1-\varepsilon)$ confidence:} We determine $p$-value for each candidate item, and if $p$-value is more than $\varepsilon$, we include it in the recommendation set; otherwise, we do not. 
\[
 \Gamma^\varepsilon = \{o_j\mid o_j \in O\setminus \{O_{V(G)}\}, p(o_j)>\varepsilon             \}
\]
\noindent The proposed conformal group recommender algorithm is outlined in Algorithm~\ref{algo:CGRS}. 
\begin{algorithm}[ht!]
		\SetAlgoLined
		\KwIn{$O, Group~ G, O_j~ \forall u_j\in G,~ Support, ~ \varepsilon$}
		\KwOut{Recommendation set ($\Gamma^{\varepsilon}$)}
		
		$O_{V(G)}\leftarrow Build-Virtual-User-Profile(Support, O_j~ \forall u_j \in G)$\; 
		
		split $O_{V(G)}$ into two sets $O^1_{V(G)}$ and
		$O^2_{V(G)}$\;
		
		$\Gamma^{\varepsilon} \leftarrow \emptyset $ \;
		
		\For{each $o'j_h$ in $O^c_j$} {
			Compute $\alpha_h$ using Equation~\ref{eq:ncm}\;
		}
		\For {each $o$ $\in O\setminus \{O_j, \forall u_j \in G\}$} {
		    $o_{n+1}\leftarrow o$\;
		    
		    $O^3_{V(G)}\leftarrow O^1_{V(G)}\cup\{o_{n+1}\}$\;
		    
		    \For{each $o'_j$ in $O^2_{V(G)}$} {
			  Compute $P(o_h|o'_j), h\in[1,n+1]$\; 
			  
			  Compute $p(o_{n+1}, o'_j)$ using Equation~\ref{eq:pval2} \;
			  
			  Compute $p(o_{n+1})$ using Equation~\ref{eq:pval3}\;
			  
		    	\lIf { $p(o_{n+1}) > \varepsilon$ } {
			    	$\Gamma^{\varepsilon} \leftarrow \Gamma^{\varepsilon} \cup \{o\}$
			    }
			}
		}
		\caption{Conformal Group Recommender Systems.}
		\label{algo:CGRS}
	\end{algorithm}

\begin{algorithm}[ht!]
		\SetAlgoLined
		\KwIn{$Support, O_j~ \forall u_j\in G$}
		\KwOut{Virtual user $V(G)$ profile $O_{V(G)}$}
		
		$O_{V(G)}\leftarrow \emptyset$\;
		
		$\tau\leftarrow input('Threshold')$\; 
		
		$Group-items \leftarrow \bigcup_{u_j\in G} O_j$\; 
		
		\For{each $o_i\in Group-items$}{
		   Compute $w_i = weight(o_i, G)$ using Equation~\ref{eq:weight}\;
		   
		   \lIf{$w_i>\tau$}{
		     $O_{V(G)}\leftarrow O_{V(G)}\cup \{o_i:w_i\}$
		   }
		}
			\caption{Build-Virtual-User-Profile}
		\label{algo:VU}
\end{algorithm}

\subsection{Validity and Efficiency}
\label{sec:valEff}
The property of \textit{validity} ensures that the probability of committing an error will not surpass $\varepsilon$ for the recommendation set $\Gamma^\varepsilon$. 
This indicates that the possibility of not recommending user's interesting item $o$  is bounded to $\varepsilon$, i.e., $P(p(o\le \varepsilon))\le \varepsilon$.
Lemma~\ref{lem:invariant} discussed in the previous subsection forms the basis for proving the validity of the proposed approach. The following lemma establishes the fulfillment of the validity property by the proposed CGRS by employing the line of reasoning provided by Vovk et al.~\cite{shaf}. 
\begin{Lemma}
		\label{lem:validity}
  Assuming that objects $o_{1}, o_{2}, \ldots, o_{n+1}$ are distributed independently and identically concerning their precedence relations with the training data, the probability of the error that $o_{n+1}$ does not belong to $\Gamma^\varepsilon$ will be at most $\varepsilon \in [0,1]$, i.e., $P(p(o_{n+1})\le \varepsilon)\le \varepsilon$.
	\end{Lemma}%
	\begin{proof}

  When the $p$-value of a new item $o_{n+1}$ to be recommended is less than or equal to $\varepsilon$, i.e., $p(o_{n+1})\le \varepsilon$, we regard that as an error. The $p$-value is lower than $\varepsilon$ when the expression $\frac{\sum_{o'j\in O^2_{V(G)}} p(o_{n+1}, o'j)}{k} \le k$ holds. To determine the probability of $p(o_{n+1}, o'_j)\le \varepsilon$ for all $o'_j$, we consider the case where $\alpha^{n+1}(o'j)$ is among the $\lfloor \varepsilon(l+1)\rfloor$ largest elements of the set $\{\alpha^{1}(o'j), \alpha^{2}(o'j), \ldots, \alpha^{n+1}(o'j)\}$.
When all objects ($\{o_1, o_2, \ldots, o_{n+1}\}$) are independently and identically distributed with respect to their precedence relations with $o'j$, all permutations of the set $\{\alpha^{1}(o'j), \alpha^{2}(o'j), \ldots, \alpha^{n+1}(o'j)\}$ are equiprobable. Thus, the probability that $\alpha^{n+1}$ is among the $\lfloor\varepsilon(l+1)\rfloor$ largest elements does not exceed $\varepsilon$. Consequently, $P( p(o_{n+1}, o'_j)\le \varepsilon)\le \varepsilon$.
In the scenario where the $p$-values, i.e., $p(o_{n+1}, o'j)$, are uniformly distributed, the mean of these $p$-values, denoted as $p(o_{n+1})$, will not exceed $\varepsilon$. Hence, $P(p(o_{n+1})\le\varepsilon)\le \varepsilon$, which represents the probability of error.
	\end{proof}
	
 Apart from meeting the validity property, possessing an efficient recommendation set is desirable. Within the conformal framework, an efficient set is characterized by narrower intervals and higher confidence levels. We experimentally investigate the efficiency aspects in Section~\ref{sec:Exp} using standard performance metrics.

\subsection{Time complexity analysis}
\label{sec:tca}
 In this section, we analyze the time complexity of the proposed approach. The time complexity of creating the virtual user profile is $O(nitem\times n_i\times\lvert G \rvert)$, where $nitem$ is the total number of items, $\lvert G \rvert$ is the group size, and $n_i$ is the number of items in each group member\footnote{If each member has consumed a different number of items, $n_i$ in the time complexity expression will be maximum among all the individual profile sizes.}.  Assuming $(n+k)$ items in the virtual user profile,  $n$ items in $O^1_{V(G)}$ and k items in $O^2_{V(G)}$, the time complexity of the $p$-value computation for each item is $O(nk)$. To make a recommendation set, we need to compute $p$-value for all the candidate items. Hence, the total complexity is $O(n\times k\times nitem + nitem\times n_i\times\lvert G \rvert). $

\section{Emperical Study}
\label{sec:Exp}
This section reports the empirical assessment of our proposed conformal group recommender system (CGRS). Experiments were carried out on eight benchmark datasets demonstrating that CGRS could achieve higher prediction accuracy than the baseline GRS model~\cite{kagita15}. Table~\ref{datasets} presents the details of the experiments used for experiments.

\begin{table}[ht!]
\centering
\caption{Experimental Datasets Description.}
\label{datasets}
\begin{tabular}{ |l|r|r|r| } 
 \hline
 \textbf{Dataset} & \textbf{Users} & \textbf{Items} & \textbf{Records} \\ 
 \hline
 MovieLens 100K & 943 & 1,682 & 100,000 \\ 
 \hline
  MovieLens-latest-small & 707 & 8,553 & 100,024 \\ 
 \hline
 MovieLens 1M & 6,040 & 3,952 & 1,000,209 \\ 
 \hline
 Personality-2018 & 1,820 & 35,196 & 1,028,752 \\ 
 \hline
 MovieLens 10M & 71,567 & 10,681 & 10,000,054 \\ 
 \hline
 MovieLens 20M & 138,494 & 26,745 & 20,000,262 \\ 
  \hline
 MovieLens-latest & 229,061 & 26,780 & 21,063,128 \\ 
 \hline
 MovieLens 25M & 162,000 & 62,000 & 25,000,096 \\ 
 \hline
\end{tabular}
\end{table}

\noindent 
In the preprocessing stage, the items available in a user's profile are ordered according to their timestamp. We remove the user profiles having fewer than 20 ratings and perform train and test division for each user in a 6:4 ratio. The virtual user profile is constructed using the data available in the training sets.  Further, in the case of CGRS, for splitting the virtual user profile into two groups, we fine-tuned with different combinations of sizes and selected 75\% of the virtual user profile as $O_{V(G)}^1$ and the remaining as the $O_{V(G)}^2$. We analyzed the results in two different kinds of group settings, namely \textit{homogeneous} and \textit{random}.  In the homogeneous group setting, the candidates for forming a group are chosen in a way such that they jointly share at least some percentage of common items. Given the size of group $g$, we form a homogeneous group if each user has at least $100/(2*g)$ percent of common items.
%
In the random group setting, the candidates for forming a group are chosen randomly from the list of users. \\



\noindent \textbf{Performance Metrics}: To assess the effectiveness of the proposed CGRS approach, we used the following evaluation metrics in our experiments \\ 
\noindent \textit{Precision:}  Precision represents the proportion of items recommended by a system that are relevant to the user's interests. 
\[
Precision = \frac{\lvert recommended \cap used\rvert }{\lvert recommended\rvert},
\]
where $recommended$ is the set of all items recommended by the system, and $used$ is the set of relevant items that the user used or interacted with. \\

\noindent \textit{Recall:} Recall measures the proportion of relevant items that were actually recommended out of all the relevant items that exist for the user.
\[Recall = \frac{\lvert recommended \cap used\rvert}{\lvert used \lvert}.\] 

\noindent
\textit{F1-score:} The F1-score is defined as the harmonic mean of precision and recall, and ranges from 0 to 1, where a higher score indicates a better performance.
\[ F1= 2 \times \frac{precision \times recall}{precision+recall}.\] 

\noindent
\textit{Normalized Discounted Cumulative Gain (NDCG):} NDCG is a widely used ranking quality metric that measures a recommendation system's effectiveness based on the recommended items' graded relevance. NDCG calculates the ratio of the Discounted Cumulative Gain (DCG) of the recommended order to the Ideal DCG (iDCG) of the order. DCG measures the relevance and position of the recommended items, while iDCG represents the highest possible DCG that can be obtained for a given set of relevant items.
\[NDCG = \frac{DCG}{iDCG},\]
where DCG is calculated as 
\[DCG = \sum_{i=1}^{n} \frac{2^{rel_i}-1}{\log_2(i+1)}\]
and iDCG is calculated by sorting the set of relevant items in decreasing order of their relevance scores.

\noindent \textit{Reverse Reciprocal (RR): } RR calculates the reciprocal rank of the first item in the recommendation set that the user actually consumes. The reciprocal rank of a user is defined as the relevance score of the top-most relevant item. The RR score is then calculated by taking the reciprocal of the rank of the first item in the recommendation set. A higher RR score indicates a higher ranking quality of the recommended items.
\[RR = \frac{1}{r}.\] 

\noindent \textit{Average Precision (AP): } AP measures the quality of ranked lists. It is calculated as the average of precision values at each point in the ranked list where a relevant item is found. Given N number of items in the recommendation set and \textit{m} number of items actually consumed by the user, the average precision score is calculated as follows
\[AP = \frac{1}{m}\sum_{k=1}^{N}P(k)\times rel(k),\]
where \textit{P(k)} is the precision at \textit{top-k }items in recommendation set. The value of \textit{rel(k)} is 1 if the item at kth position in the recommendation set is actually consumed by the user, otherwise 0.

\noindent \textit{Area Under Curve (AUC): }
AUC measures the probability that a random relevant item will be ranked higher than a random irrelevant item.  The higher value of AUC score implies a better recommendation system. AUC ranges between 0 and 1, with a higher value indicating a better ranking performance of the recommendation system.

\subsection{Experimental Results}
This section reports the experimental analysis to study the performance of CGRS. The experiments are conducted on standard benchmark datasets mentioned in Table~\ref{datasets}. The evaluation is based on traditional metrics such as Precision, Recall, F1-score, NDCG, RR, AP, and AUC.  We also analyze the proposed approach by varying the group sizes. All the results reported here are the average of 500 randomly generated instances.



\subsubsection{Performance over different datasets}
We evaluated the Conformal Group Recommender System (CGRS) against the Group Recommender System (GRS) [7], using
the performance metrics mentioned earlier in this section. GRS [7] serves as the foundation for our proposed CGRS. We conducted experiments using various datasets listed in Table~\ref{datasets}, and the results indicate that the conformal approach improves recommendation accuracy and reliability by providing confidence. Although we present the results for a group size of 2, we observed similar outcomes for other group sizes.\\
  
\noindent \textit{Homogeneous Groups: } In the first set of experiments, we assessed the effectiveness of our
proposed approach in the homogeneous group setting. The performance results for the homogeneous group setting are presented in Table~\ref{tab:homo-perf}.

\noindent \textit{}
\begin{table}[H]
\small
\centering
\scalebox{0.98}{
\begin{tabular}{|c|c|c|c|c|c|c|c|c|} 
\hline
 & \multicolumn{2}{c|}{\textbf{AP}} 
& \multicolumn{2}{c|}{\textbf{RR}} 
 & \multicolumn{2}{c|}{\textbf{AUC}} 
  & \multicolumn{2}{c|}{\textbf{NDCG}} \\ 
\hline
\textbf{Dataset}   & \textbf{GRS} & \textbf{CGRS}  & \textbf{GRS} & \textbf{CGRS}  & \textbf{GRS} & \textbf{CGRS}  & \textbf{GRS} & \textbf{CGRS}  \\ 
\hline
\textbf{ML 100K} & 	 0.20563	& \textbf{0.22068} &	0.74760	& \textbf{0.74972} &  0.88378	& \textbf{0.89307} &	0.56149	& \textbf{0.57118} \\ \hline
\textbf{ML-LS} & 	 0.18439	& \textbf{0.25334} &	0.75768	& \textbf{0.80948} &  0.94065	& \textbf{0.95401} &	0.55016	& \textbf{0.58754} \\ \hline
\textbf{ML 1M} & 	 0.20712	& \textbf{0.21363} &	0.81423	& \textbf{0.81464} &  0.88204	& \textbf{0.88616} &	0.60240	& \textbf{0.60578} \\ \hline
\textbf{ML 10M} &   0.21730	& \textbf{0.21929} &	0.74704	& \textbf{0.74891}  & 0.97131	& \textbf{0.97307} &	0.55389	& \textbf{0.55569}  \\ \hline
\textbf{ML 20M} &   0.19713 & \textbf{0.19925} &	0.70836	& \textbf{0.70961}  & 0.98494	& \textbf{0.98589} &	0.54107	& \textbf{0.54267}  \\ \hline
\textbf{ML 25M} &   0.18300	& \textbf{0.18455} &	0.75612	& 0.75612  & 0.99177	& \textbf{0.99231} &	0.54850 & \textbf{0.54979}  \\ \hline
\textbf{ML Latest} &   0.19361	& \textbf{0.19494} &	0.66250	& \textbf{0.66250}  & 0.98810	& \textbf{0.98850} &	0.52236 & \textbf{0.52341}  \\ \hline
\textbf{Per-2018} &   0.21536	& \textbf{0.27154} &	0.89691	& \textbf{0.90925}  & 0.96067	& \textbf{0.96489} &	 0.63456 & \textbf{0.65580}  \\ \hline
\end{tabular}
}
\caption{Performance comparison between GRS and CGRS for different datasets for homogenous group setting. The proposed CGRS outperforms GRS in terms of AP, RR, AUC and NDCG for all the datasets.}\label{tab:homo-perf}
\end{table}

\begin{figure}[H]
    \centering
    \includegraphics[width=3.8cm]{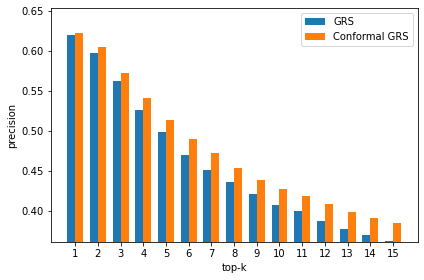}
    \includegraphics[width=3.8cm]{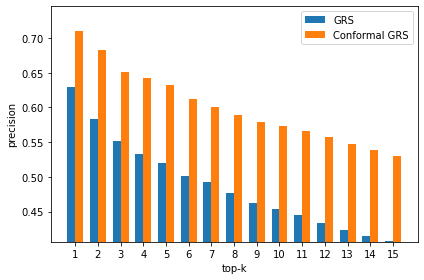}
    \includegraphics[width=3.8cm]{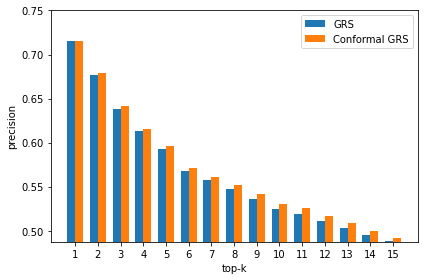}
    \includegraphics[width=3.8cm]{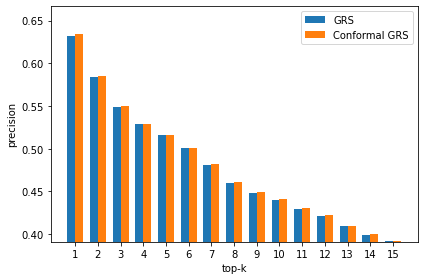}
    \includegraphics[width=3.8cm]{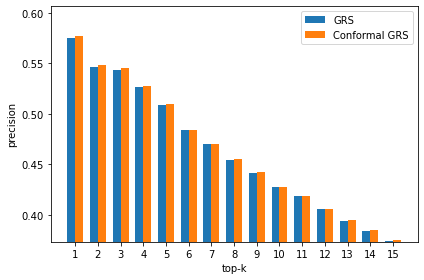}
    \includegraphics[width=3.8cm]{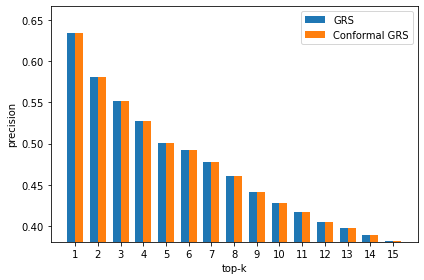}
    \includegraphics[width=3.8cm]{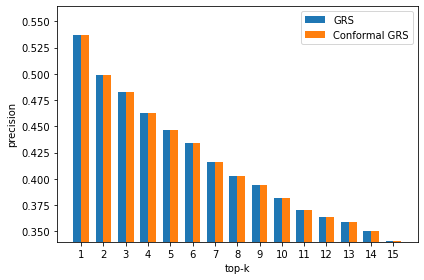}
    \includegraphics[width=3.8cm]{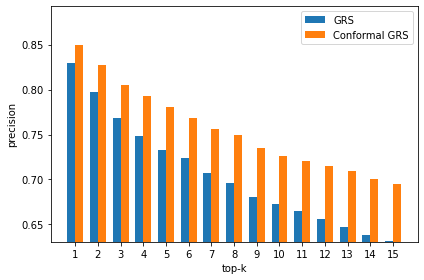}
    \caption{Comparision of Precision between CGRS and GRS for MovieLens 100K, MovieLens-latest-small, MovieLens 1M, MovieLens 10M, MovieLens 20M, MovieLens 25M, MovieLens latest, Personality 2018 for homogeneous groups. The figures are arranged from left to right and top to bottom.}
    \label{fig:prec-homo-datasets}
\end{figure}

\begin{figure}[H]
    \centering
    \includegraphics[width=3.8cm]{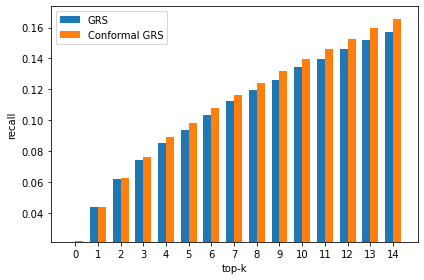}
    \includegraphics[width=3.8cm]{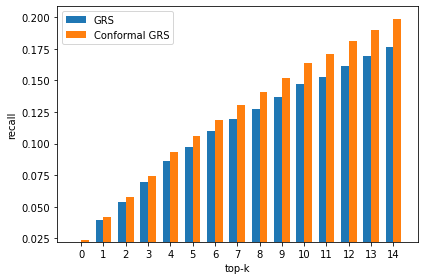}
    \includegraphics[width=3.8cm]{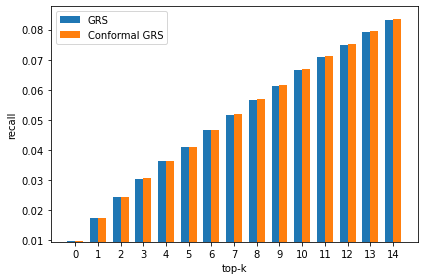}
    \includegraphics[width=3.8cm]{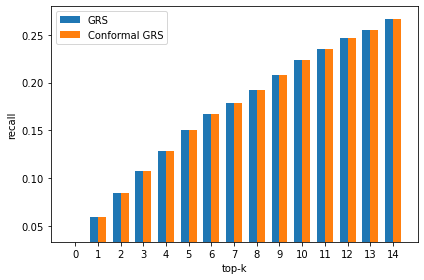}
    \includegraphics[width=3.8cm]{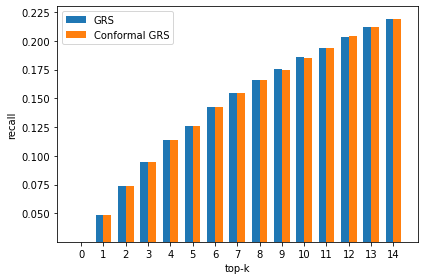}
    \includegraphics[width=3.8cm]{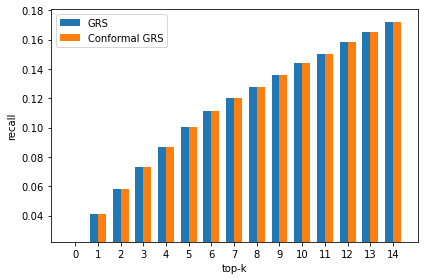}
    \includegraphics[width=3.8cm]{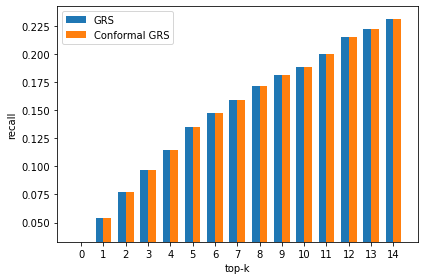}
    \includegraphics[width=3.8cm]{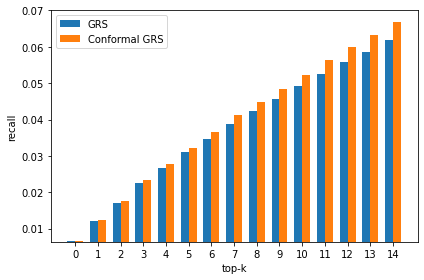}
    \caption{Comparision of Recall between CGRS and GRS for MovieLens 100K, MovieLens-latest-small, MovieLens 1M, MovieLens 10M, MovieLens 20M, MovieLens 25M, MovieLens latest, Personality 2018 for homogeneous groups. The figures are arranged from left to right and top to bottom.}
    \label{fig:recall-homo-datasets}
\end{figure}

\begin{figure}[H]
    \centering
    \includegraphics[width=3.8cm]{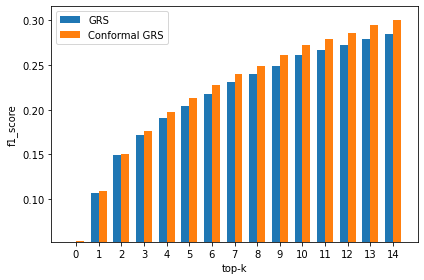}
    \includegraphics[width=3.8cm]{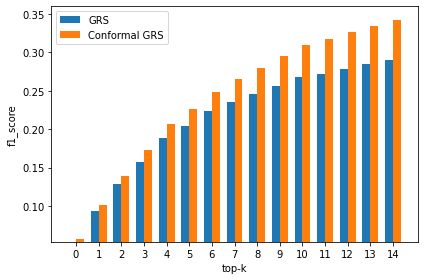}
    \includegraphics[width=3.8cm]{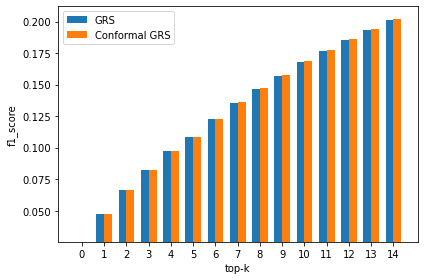}
    \includegraphics[width=3.8cm]{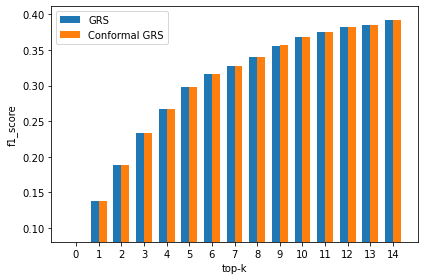}
    \includegraphics[width=3.8cm]{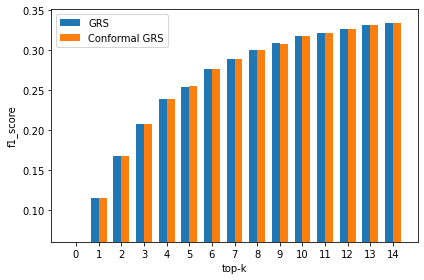}
    \includegraphics[width=3.8cm]{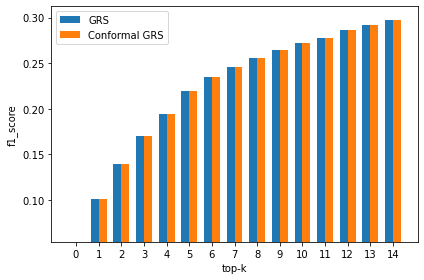}
    \includegraphics[width=3.8cm]{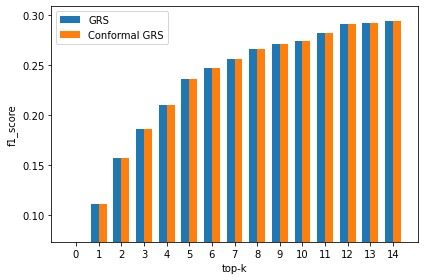}
    \includegraphics[width=3.8cm]{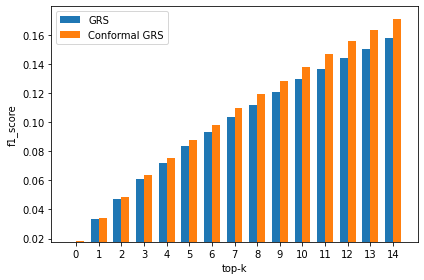}
    \caption{Comparision of Recall between CGRS and GRS for MovieLens 100K, MovieLens-latest-small, MovieLens 1M, MovieLens 10M, MovieLens 20M, MovieLens 25M, MovieLens latest, Personality 2018 for homogeneous groups. The figures are arranged from left to right and top to bottom.}
    \label{fig:f1-homo-datasets}
\end{figure}

In Figure~\ref{fig:prec-homo-datasets}, we present a comprehensive comparison of precision scores for different datasets, while varying the top-K parameter. The results indicate that our proposed CGRS outperforms the underlying GRS in terms of precision scores. It is noteworthy that we obtained similar results for recall and F1-score as well, which are presented in Figures~\ref{fig:recall-homo-datasets} and \ref{fig:f1-homo-datasets}, respectively. These findings demonstrate that the proposed CGRS can deliver better recommendation accuracy compared to the traditional GRS, and can be particularly useful in applications where precise recommendations are crucial.

\begin{figure}[H]
    \centering
    \includegraphics[width=3.8cm]{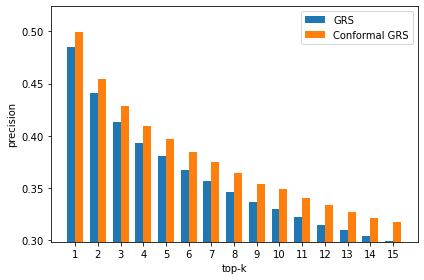}
    \includegraphics[width=3.8cm]{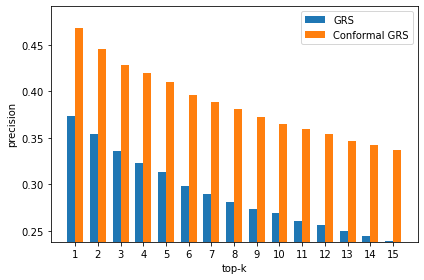}
    \includegraphics[width=3.8cm]{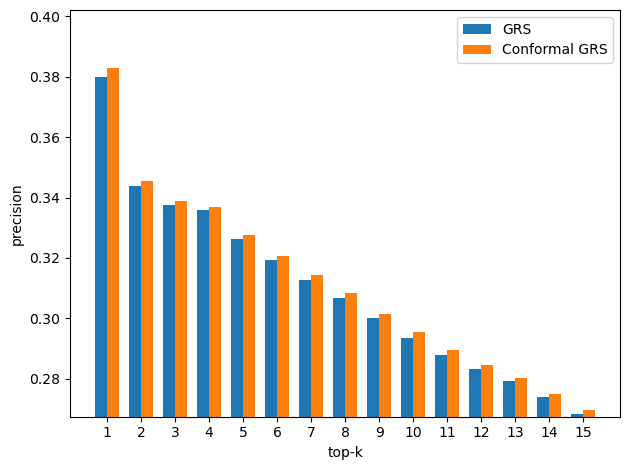}
    \includegraphics[width=3.8cm]{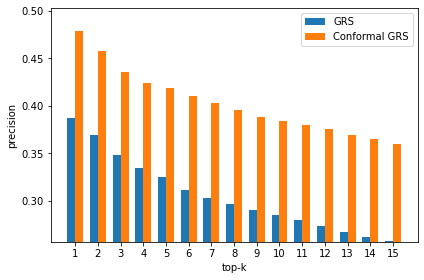}
    \includegraphics[width=3.8cm]{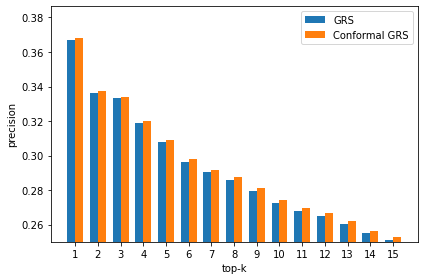}
    \includegraphics[width=3.8cm]{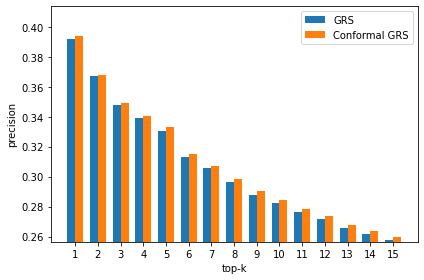}
    \includegraphics[width=3.8cm]{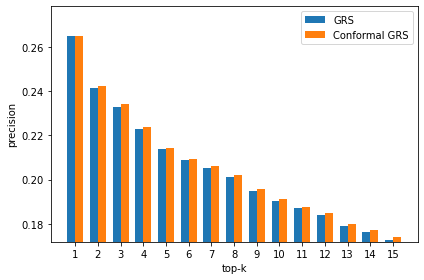}
    \includegraphics[width=3.8cm]{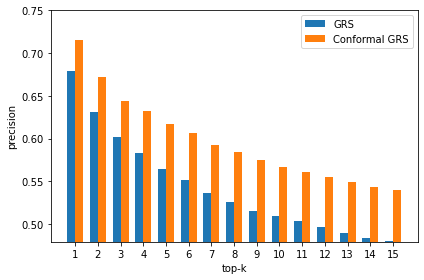}
    \caption{Comparision of Precision between CGRS and GRS for MovieLens 100K, MovieLens-latest-small, MovieLens 1M, MovieLens 10M, MovieLens 20M, MovieLens 25M, MovieLens latest, Personality 2018 for random groups. The figures are arranged from left to right and top to bottom.}
    \label{fig:prec-non-homo-datasets}
\end{figure}

\begin{figure}[H]
    \centering
    \includegraphics[width=3.8cm]{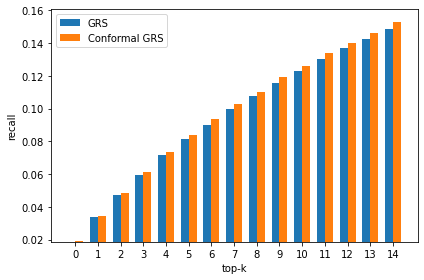}
    \includegraphics[width=3.8cm]{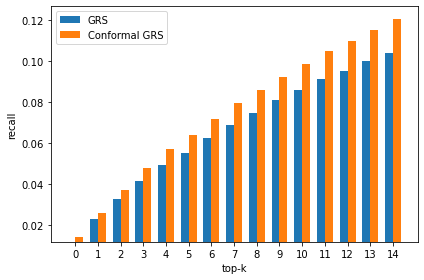}
    \includegraphics[width=3.8cm]{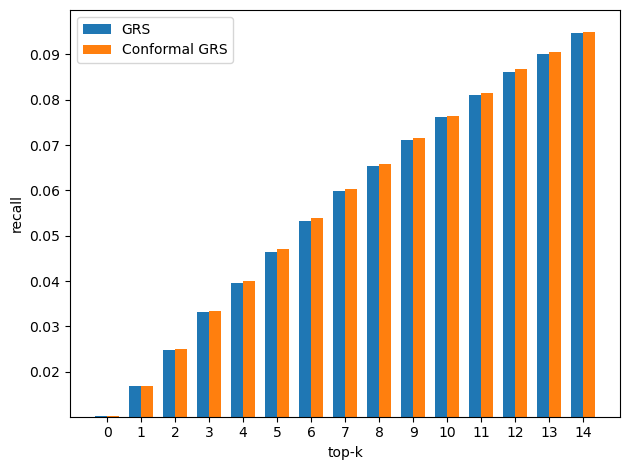}
    \includegraphics[width=3.8cm]{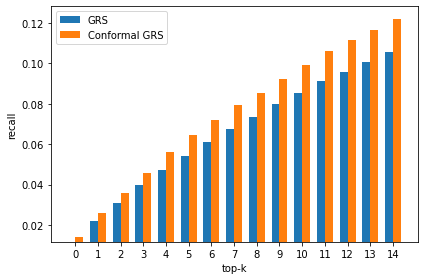}
    \includegraphics[width=3.8cm]{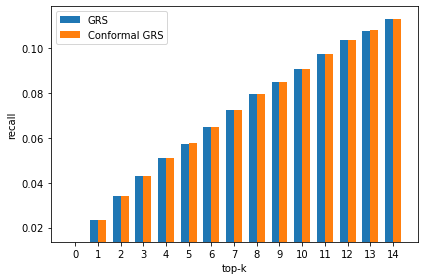}
    \includegraphics[width=3.8cm]{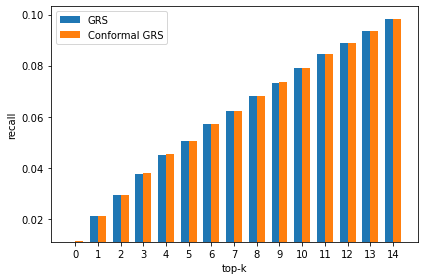}
    \includegraphics[width=3.8cm]{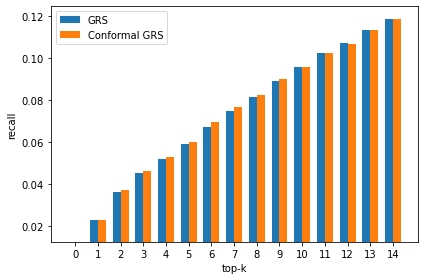}
    \includegraphics[width=3.8cm]{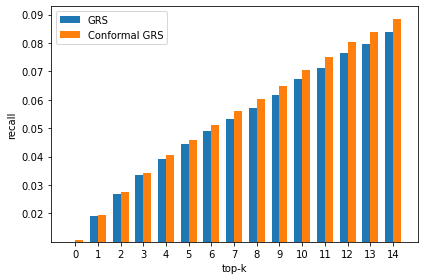}
    \caption{Comparision of Recall between CGRS and GRS for MovieLens 100K, MovieLens-latest-small, MovieLens 1M, MovieLens 10M, MovieLens 20M, MovieLens 25M, MovieLens latest, Personality 2018 for random groups. The figures are arranged from left to right and top to bottom.}
    \label{fig:recall-non-homo-datasets}
\end{figure}

\begin{figure}[H]
    \centering
    \includegraphics[width=3.8cm]{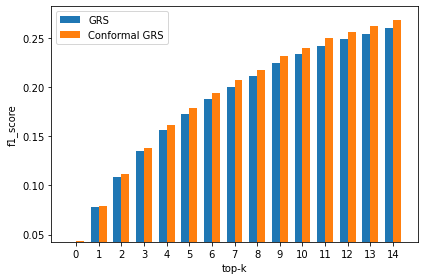}
    \includegraphics[width=3.8cm]{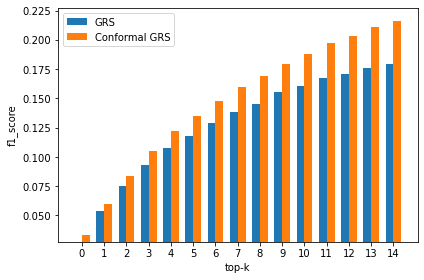}
    \includegraphics[width=3.8cm]{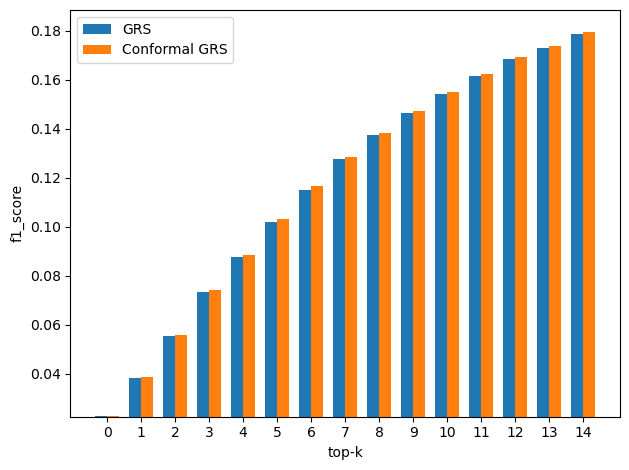}
    \includegraphics[width=3.8cm]{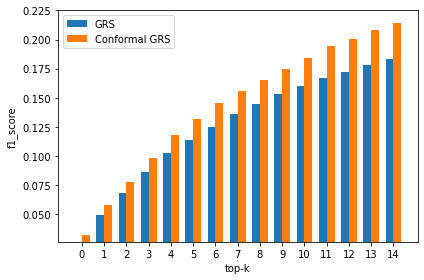}
    \includegraphics[width=3.8cm]{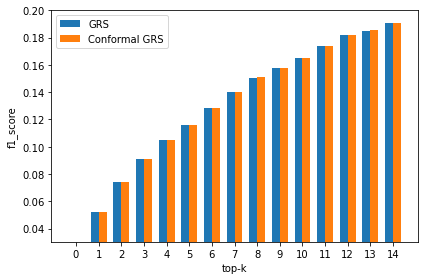}
    \includegraphics[width=3.8cm]{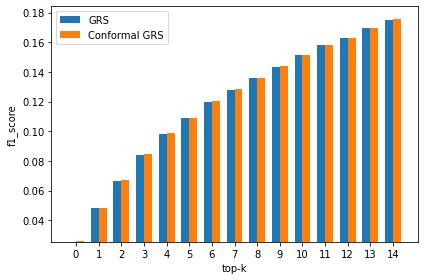}
    \includegraphics[width=3.8cm]{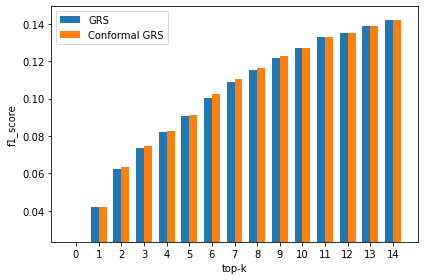}
    \includegraphics[width=3.8cm]{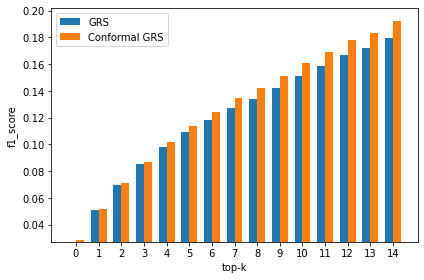}
    \caption{Comparision of Recall between CGRS and GRS for MovieLens 100K, MovieLens-latest-small, MovieLens 1M, MovieLens 10M, MovieLens 20M, MovieLens 25M, MovieLens latest, Personality 2018 for random groups. The figures are arranged from left to right and top to bottom.}
    \label{fig:f1-non-homo-datasets}
\end{figure}

\noindent \textit{Random Groups: } In this experimental section, we focus on the performance results for the random group setting. Table~\ref{tab:non-homo-perf} displays the AP, RR, AUC, and NDCG values of the CGRS and GRS. Our findings indicate that the CGRS approach continues to outperform the GRS approach, and we observe a similar trend in the random group setting. Additionally, we have included visual representations of the Precision, Recall, and F1-score metrics for each dataset in the Figures~\ref{fig:prec-non-homo-datasets}, \ref{fig:recall-non-homo-datasets}, and \ref{fig:f1-non-homo-datasets}, respectively.

\noindent \textit{}
\begin{table}[H]
\small
\centering
\scalebox{0.97}{
\begin{tabular}{|c|c|c|c|c|c|c|c|c|} 
\hline
 & \multicolumn{2}{c|}{\textbf{AP}} 
& \multicolumn{2}{c|}{\textbf{RR}} 
 & \multicolumn{2}{c|}{\textbf{AUC}} 
  & \multicolumn{2}{c|}{\textbf{NDCG}} \\ 
\hline
\textbf{Dataset}   & \textbf{GRS} & \textbf{CGRS}  & \textbf{GRS} & \textbf{CGRS}  & \textbf{GRS} & \textbf{CGRS}  & \textbf{GRS} & \textbf{CGRS}  \\ 
\hline
\textbf{ML 100K} & 	 0.18899	& \textbf{0.20271} &	0.62167	& \textbf{0.63175} &  0.87293	& \textbf{0.87784} &	0.54847	& \textbf{0.55597} \\ \hline
\textbf{ML-LS} & 	 0.12772	& \textbf{0.19950} &	0.51417	& \textbf{0.57387} &  0.91507	& \textbf{0.92548} &	0.50005	& \textbf{0.53692} \\ \hline
\textbf{ML 1M} & 	  0.14462	& \textbf{0.14849} &	0.52104	& \textbf{0.52397} &  0.87487	& \textbf{0.87733} &	0.52774	& \textbf{0.53016} \\ \hline
\textbf{ML 10M} &   0.12972 & \textbf{0.20035} &	0.52734	& \textbf{0.58781}  & 0.91666	& \textbf{0.92339} &	0.50428	& \textbf{0.54104}  \\ \hline
\textbf{ML 20M} &   0.14143 & 0.14143 &	0.50034	& \textbf{0.50107}  & 0.97454	& \textbf{0.97538} &	0.50591	& \textbf{0.50745}  \\ \hline
\textbf{ML 25M} &   0.13341	& \textbf{0.13665} &	0.52560	& \textbf{0.52754} & 0.98546 & \textbf{0.98603} &	0.51164 & \textbf{0.51330}  \\ \hline
\textbf{ML Latest} &   0.10681 & \textbf{0.10915} &	0.36672	& \textbf{0.36697}  & 0.98266	& \textbf{0.98297} &	0.41387 & \textbf{0.41514}  \\ \hline
\textbf{Per-2018} &   0.18746	& \textbf{0.23878} &	0.77307	& \textbf{0.79475}  & 0.95812	& \textbf{0.96218} &	0.59667 & \textbf{0.61593}  \\ \hline
\end{tabular}
}
\caption{Performance comparsion between GRS and CGRS for different datasets for random
group setting. The proposed CGRS outperforms GRS in terms of AP, RR, AUC and NDCG for all the datasets.}\label{tab:non-homo-perf}
\end{table}

\subsubsection{Effect of Varying Group Sizes on Performance of CGRS } This subsection presents the comparative performance analysis of CGRS and GRS for homogeneous and random groups with varying group sizes. The CGRS method consistently outperforms GRS, as shown in the previous section. 

\noindent \textit{Homogeneous Groups: }
The results in Table~\ref{tab:group-size-perf-homo} show the performance metrics values for the varying group size. We present results specific to the Personality-2018 dataset in the table. Similar outcomes have been observed over the other datasets.


\begin{table}[H]
\small
\centering
\scalebox{0.96}{
\begin{tabular}{|c|c|c|c|c|c|c|c|c|} 
\hline
 & \multicolumn{2}{c|}{\textbf{AP}} 
& \multicolumn{2}{c|}{\textbf{RR}} 
 & \multicolumn{2}{c|}{\textbf{AUC}} 
  & \multicolumn{2}{c|}{\textbf{NDCG}} \\ 
\hline
\textbf{Group size}   & \textbf{GRS} & \textbf{CGRS}  & \textbf{GRS} & \textbf{CGRS}  & \textbf{GRS} & \textbf{CGRS}  & \textbf{GRS} & \textbf{CGRS}  \\ 
\hline
\textbf{2} & 0.21536      & \textbf{0.27154}       & 0.89691     & \textbf{0.90925}    & 0.96067     & \textbf{0.96489} & 0.63456       & \textbf{0.65580}             \\ 
\hline
\textbf{3} & 0.15776      & \textbf{0.19241}       & 0.90760     & \textbf{0.92286}    & 0.95538     & \textbf{0.95840} & 0.52325      & \textbf{0.53786}             \\ 
\hline
\textbf{4} & 0.10855      & \textbf{0.13315}       & 0.87872     & \textbf{0.88972}    & 0.95073     & \textbf{0.95301} & 0.43393      & \textbf{0.44459}             \\ 
\hline
\textbf{5} & 0.07871      & \textbf{0.09692}       & 0.80740     & \textbf{0.82353}   & 0.94608     & \textbf{0.94773}   & 0.36742      & \textbf{0.37538}           \\ 
\hline
\textbf{6} & 0.06259      & \textbf{0.07651}       & 0.78733     & \textbf{0.81640}     & 0.94102     & \textbf{0.94229} & 0.32410      & \textbf{0.33045 }          \\
\hline
\end{tabular}
}
\caption{\label{tab:group-size-perf-homo}Performance comparsion between GRS and CGRS for Personality-2018 dataset for homogenous group setting. The proposed CGRS method shows better results in terms of AP, RR, AUC and NDCG for all the group sizes. }
\end{table}

\begin{figure}[H]
    \label{fig:prec1}
    \centering
    \includegraphics[width=5.4cm]{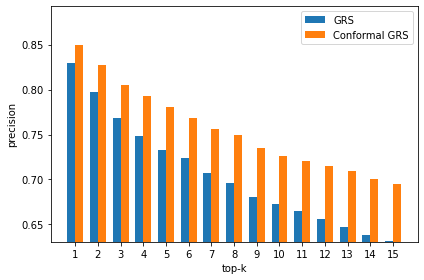}
    \includegraphics[width=5.4cm]{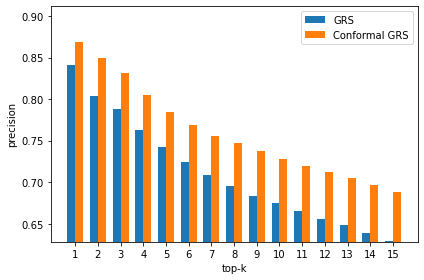}
    \includegraphics[width=5.4cm]{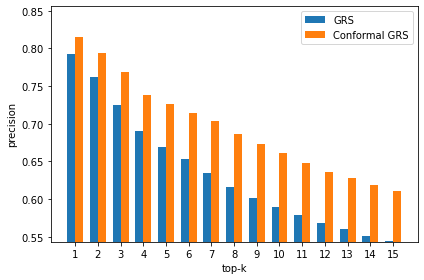}
    \includegraphics[width=5.4cm]{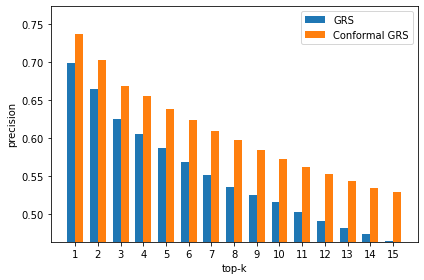}
    \includegraphics[width=5.4cm]{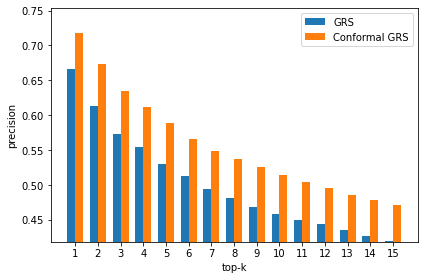}
    \caption{Comparison of Precision between the conformal and base method of the recommender system for varying homogeneous group sizes, including group sizes 2, 3, 4, 5, and 6. The figures are arranged from left to right and top to bottom.}
    \label{fig:homo-prec6}
\end{figure}

\begin{figure}[H]
    \label{fig:Recall1}

    \centering
    \includegraphics[width=5.4cm]{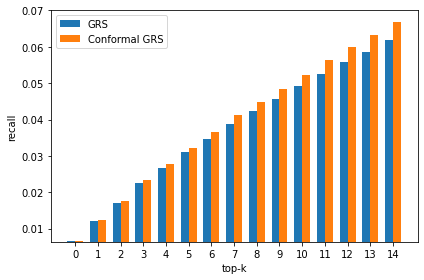}
    \includegraphics[width=5.4cm]{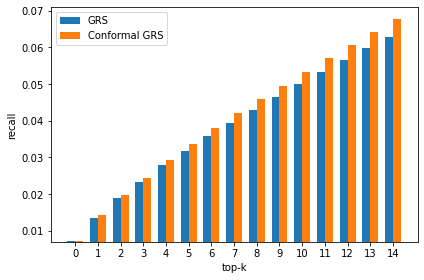}
    \includegraphics[width=5.4cm]{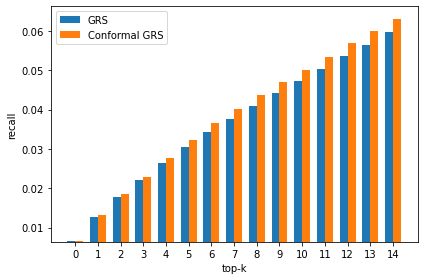}
    \includegraphics[width=5.4cm]{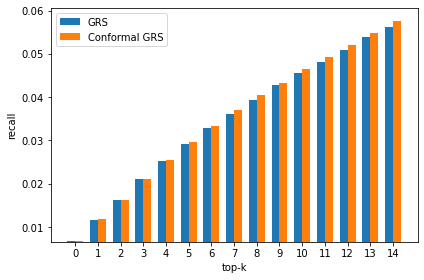}
    \includegraphics[width=5.4cm]{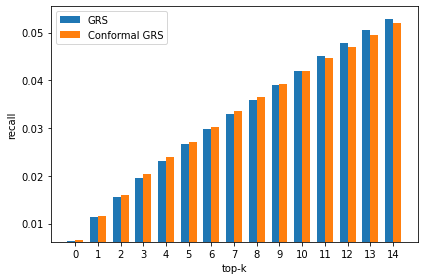}
    \caption{Comparison of Recall between the conformal and base method of the recommender system for varying homogeneous group sizes, including group sizes 2, 3, 4, 5, and 6. The figures are arranged from left to right and top to bottom.}
    \label{fig:homo-rec6}
\end{figure}

\begin{figure}[H]
    \label{fig:f11}
    \centering
    \includegraphics[width=5.4cm]{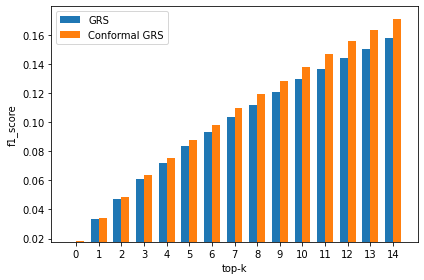}
    \includegraphics[width=5.4cm]{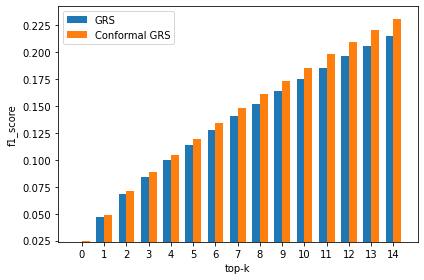}
    \includegraphics[width=5.4cm]{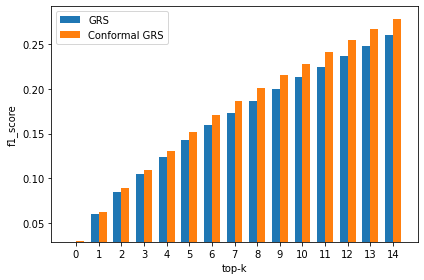}
    \includegraphics[width=5.4cm]{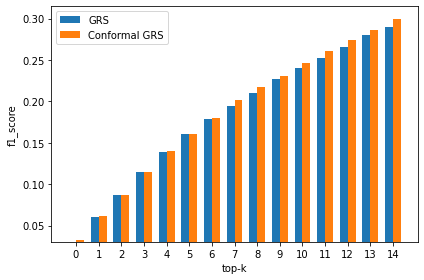}
    \includegraphics[width=5.4cm]{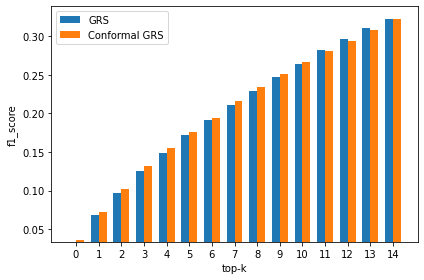}
    \caption{Comparison of F1-score between the conformal and base method of the recommender system for varying homogeneous group sizes, including group sizes 2, 3, 4, 5, and 6. The figures are arranged from left to right and top to bottom.}
    \label{fig:homo-f16}

\end{figure}

We also provide visual comparisons of the Precision, Recall, and F1-score metrics for different group sizes using Figures~\ref{fig:homo-prec6}, \ref{fig:homo-rec6} and \ref{fig:homo-f16}, respectively. Our results show that the proposed CGRS method outperforms GRS in terms of precision, recall, and F1-score for homogeneous groups with various group sizes. The improvement is particularly noticeable in the precision metric.

\vspace{0.5cm}

\vspace{0.5cm}
\vspace{0.5cm}
\noindent \textit{Random groups: }
In this series of experiments, we investigate the performance of the proposed CGRS compared to the base method for random groups by varying the group sizes. Table~\ref{tab:group-size-perf-nonhomo} shows the results we obtained for the Personality-2018 dataset. Our findings indicate that the CGRS approach continues to outperform the GRS approach, and we observe a similar trend in the random group setting.

\begin{table}[H]
\small
\centering
\scalebox{0.96}{
\begin{tabular}{|c|c|c|c|c|c|c|c|c|} 
\hline
 & \multicolumn{2}{c|}{\textbf{AP}} 
& \multicolumn{2}{c|}{\textbf{RR}} 
 & \multicolumn{2}{c|}{\textbf{AUC}} 
  & \multicolumn{2}{c|}{\textbf{NDCG}} \\ 
\hline
\textbf{Group size}   & \textbf{GRS} & \textbf{CGRS}  & \textbf{GRS} & \textbf{CGRS}  & \textbf{GRS} & \textbf{CGRS}  & \textbf{GRS} & \textbf{CGRS}  \\ 
\hline
\textbf{2} & 	0.18746	& \textbf{0.23878} &	0.77307	& \textbf{0.79475} &  0.95811	& \textbf{0.96218} &	0.59667	& \textbf{0.61593} \\ \hline
\textbf{3} &	0.14278	& \textbf{0.17449} &	0.70996	& \textbf{0.73890} &	0.95421	& \textbf{0.95736} &	0.51699	& \textbf{0.52834} \\ \hline
\textbf{4} &	0.11274	& \textbf{0.13342} &	0.65856	& \textbf{0.67120} &	0.94705	& \textbf{0.94906} &	0.44609	& \textbf{0.4519} \\ \hline
\textbf{5} &	0.09147	& \textbf{0.11074} &	0.63788	& \textbf{0.65718} &	0.94661	& \textbf{0.94793} &	0.40132 & \textbf{0.40617} \\ \hline
\textbf{6} &	0.07448	& \textbf{0.08751} &	0.58739	& \textbf{0.59873} &	0.93640	& \textbf{0.93714} &	0.35557 & \textbf{0.35651} \\ \hline
\end{tabular}
}
\caption{Performance comparison between GRS and CGRS for Personality-2018 dataset for the random group setting. The proposed CGRS method shows better results in terms of AP, RR, AUC and NDCG for all the group sizes.}\label{tab:group-size-perf-nonhomo}
\end{table}

\begin{figure}[H]
    \centering
    \includegraphics[width=5.4cm]{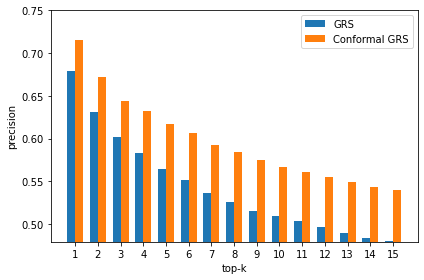}
    \includegraphics[width=5.4cm]{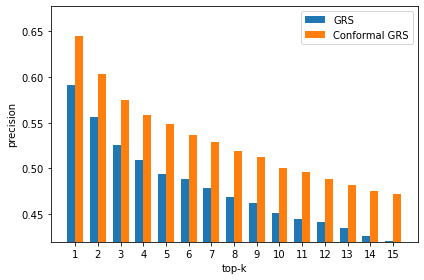}
    \includegraphics[width=5.4cm]{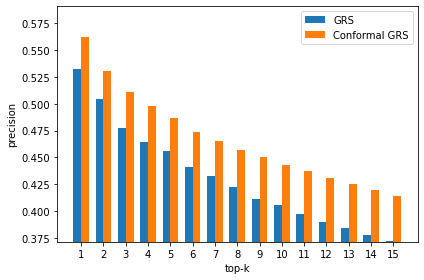}
    \includegraphics[width=5.4cm]{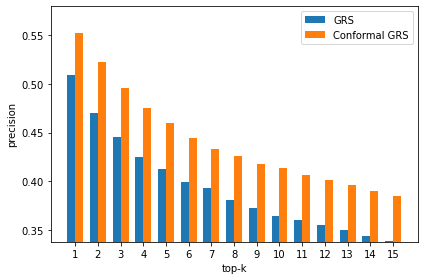}
    \includegraphics[width=5.4cm]{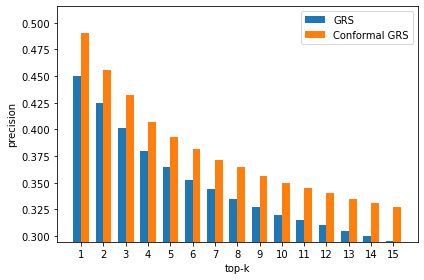}
    \caption{Comparison of Precision between the conformal and base method of the recommender system for varying random group sizes, including group sizes 2, 3, 4, 5, and 6. The figures are arranged from left to right and top to bottom.}
    \label{fig:nhomo-prec6}
\end{figure}

\begin{figure}[H]
    \centering
    \includegraphics[width=5.4cm]{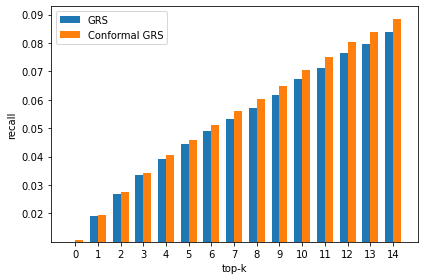}
    \includegraphics[width=5.4cm]{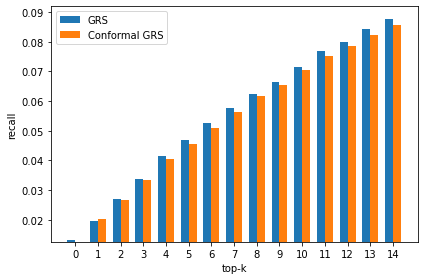}
    \includegraphics[width=5.4cm]{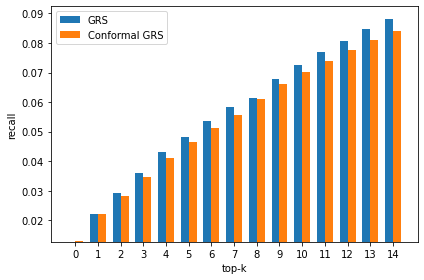}
    \includegraphics[width=5.4cm]{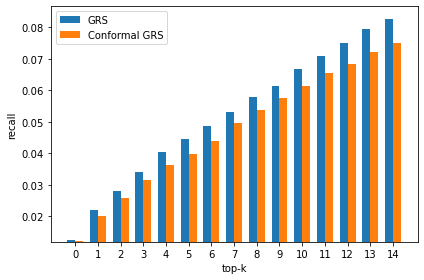}
    \includegraphics[width=5.4cm]{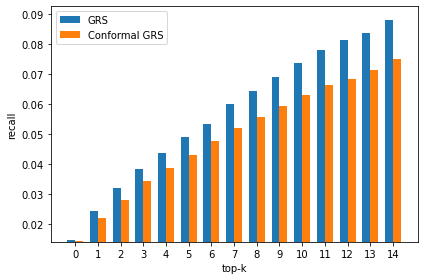}
    \caption{Comparison of Recall between the conformal and base method of the recommender system for varying random group sizes, including group sizes 2, 3, 4, 5, and 6. The figures are arranged from left to right and top to bottom.}
    \label{fig:nhomo-rec6}
\end{figure}

\begin{figure}[H]
    \centering
    \includegraphics[width=5.4cm]{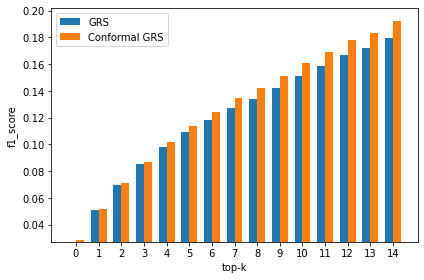}
    \includegraphics[width=5.4cm]{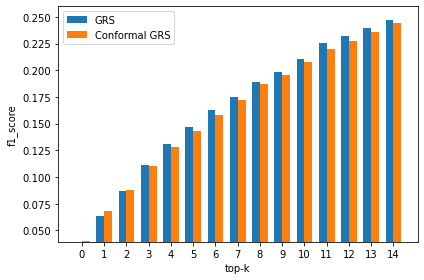}
    \includegraphics[width=5.4cm]{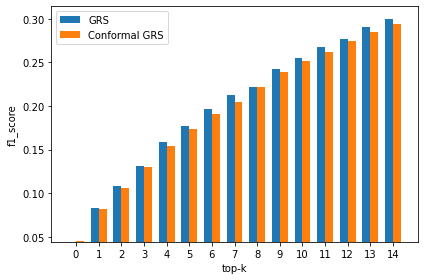}
    \includegraphics[width=5.4cm]{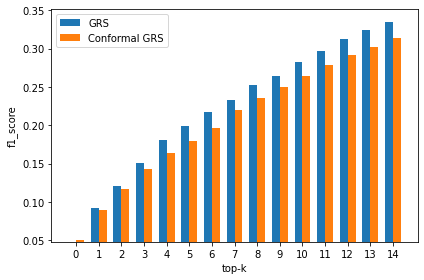}
    \includegraphics[width=5.4cm]{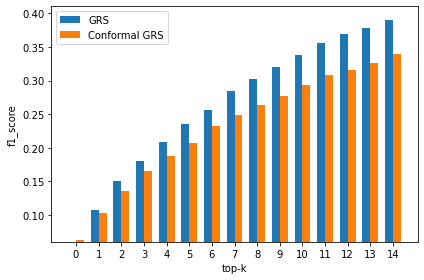}
    \caption{Comparison of F1-score between the conformal and base method of the recommender system for varying random group sizes, including group sizes 2, 3, 4, 5, and 6. The figures are arranged from left to right and top to bottom.}
    \label{fig:nhomo-f16}
\end{figure}

Figures~\ref{fig:nhomo-prec6},~\ref{fig:nhomo-rec6}, and~\ref{fig:nhomo-f16} depict a comparison between the Precision, Recall, and F1-score of the proposed CGRS and the base method for random group settings. The results indicate that the proposed CGRS method achieves better precision values compared to the base method for random groups with different group sizes. While CGRS demonstrates better recall and f1-score for random groups of size 2, GRS yields better values for other group sizes. In summary, our findings confirm the proposed CGRS achieves better performance accuracy than its counterparts while making the recommendation more transparent with the added confidence value.

\vspace{0.5cm}
\vspace{0.5cm}

\section{Conclusions and Discussion}
\label{sec:ConFw}
This paper introduces a conformal framework to the group recommendation scenario for a reliable recommendation. The theoretical facets in the article demonstrate the likelihood that the proposed CGRS makes an error is bounded by the given significance level $\varepsilon$, and hence the system exhibits a confidence of ($1-\varepsilon$). In addition to furnishing a confidence measure of reliability, the proposed method also improves the quality of recommendations. Our experimental analysis of various benchmark datasets corroborates that the proposed CGRS performs better than the baseline GRS approach in terms of different standard performance metrics assessing recommendation quality.  Extension of the proposed framework to various group recommendation algorithms is a goal worth pursuing in the future. Further, investigating a conformal framework that efficiently furnishes confidence to the complex group recommender algorithms, such as deep learning-based models, is also an exciting direction for vigorous research.


\bibliographystyle{unsrt}
\bibliography{ecai}

\end{document}